\documentclass[useAMS,usenatbib]{mn2e}

\usepackage{epsfig}
\usepackage{deluxetable}

\def\h2{H$_2$}
\def\f0{$F_0$}

\newcommand\ion[2]{#1$\;${\small\rmfamily\@Roman{#2}}\relax}%

\title[Average power density spectrum of long GRBs]{Average power density spectrum of long GRBs detected with {\em BeppoSAX}/GRBM and with {\em Fermi}/GBM}
\author[S.~Dichiara et~al.]{S.~Dichiara,$^{1}$\thanks{E-mail:dichiara@fe.infn.it}
C.~Guidorzi,$^{1}$ L.~Amati,$^{2}$ F.~Frontera$^{1,2}$ \\
\mbox{}\\
$^{1}$Department of Physics, University of Ferrara, via Saragat 1,
  I-44122, Ferrara, Italy\\
$^{2}$INAF, Istituto di Astrofisica Spaziale e Fisica Cosmica, Bologna, Via Gobetti 101,
I-40129 Bologna, Italy\\
}

\begin{document}

\date{\today}


\maketitle

\label{firstpage}

\begin{abstract}
From past experiments the average power density spectrum (PDS) of GRBs with
unknown redshift was found to be modelled from $0.01$ to 1~Hz with a
power--law, $f^{-\alpha}$, with $\alpha$ broadly consistent with $5/3$.
Recent analyses of the \textit{Swift}/BAT catalogue showed analogous results
in the 15--150 keV band.
We carried out the same analysis on the bright GRBs detected by \textit{BeppoSAX}/GRBM
and \textit{Fermi}/GBM. The \textit{BeppoSAX}/GRBM data, in the energy range 40--700 keV 
and with $7.8$ and $0.5$~ms time resolutions, allowed us to explore for the first 
time the average PDS at very high frequencies (up to 1~kHz) and reveal a break 
around 1--2~Hz, previously found in \textit{CGRO}/BATSE data.
The \textit{Fermi}/GBM data, in the energy band 8--1000~keV, allowed us to explore 
for the first time the average PDS within a broad energy range.
Our results confirm and extend the energy dependence of the PDS slope, according to
which harder photons have shallower PDS. 
\end{abstract}

\begin{keywords}
gamma-rays: bursts --- timing analysis: power density spectra
\end{keywords}

\section{Introduction}
\label{sec:intro}

Together with the energy spectrum, the temporal behaviour of gamma--ray burst
(GRB) light curves holds the key to both the physical mechanism
responsible for the production of the prompt gamma rays and the distance from the
stellar progenitor at which the energy dissipation into gamma--rays takes place.
More than a decade after the first GRB afterglow
discoveries, these key questions concerning the GRB prompt emission are yet to
be answered. The typical observed durations of pulses span from hundreds milliseconds
up to several seconds (e.g., \citealt{Norris96}).
A proper characterisation of the temporal properties at
different energy bands is crucial to provide clues to the energy dissipation process
at the origin of the gamma--rays. In this context, the average power density spectrum
(PDS) provides a way to characterise the phenomenon in terms of a stochastic process
starting from the null hypothesis that each long GRB is a different realisation of a
general unique process.
In other words, we assume that the same mechanism can explain the variability observed
in different light curves, while the observed variety is due to different conditions, which may
vary from different GRBs.

The question whether GRB light curves might entirely be explained in terms
of different realisations of a unique stochastic process characterised by a pure red
noise, is still open. Interestingly, recent analyses have found evidence for
the presence of deterministic components (as opposed to pure stochastic noise)
ruling the evolution of a GRB light curve and giving rise to a chaotic behaviour
\citep{Greco11}.

In the context of a pure stochastic process entirely characterised by red noise,
Beloborodov, Stern \& Svensson, in 1998 and 2000 \nocite{Beloborodov98}\nocite{Beloborodov00}
(hereafter, BSS98 and BSS00), studied the average PDS of 527 GRBs detected by the Burst
and Transient Source Experiment (BATSE; \citealt{Paciesas99}) aboard the
\textit{Compton Gamma Ray Observatory} (CGRO) in 25--2000 keV energy band,
revealing a typical power--law behaviour spanning almost two orders of magnitude
in frequency, from a few $10^{-2}$ to $\sim1$~Hz.
The power--law index they found is compatible with $5/3$, which is what
one expects for the Kolmogorov spectrum of velocity fluctuations within a medium
characterised by fully developed turbulence. They also found a sharp break around 1--2~Hz.
These results were also supported by the INTEGRAL data analysis of a sample of 10
bright GRB \citep{Ryde03}.

A recent analysis of the average PDS of the \textit{Swift} Burst Alert Telescope (BAT; \citealt{Barthelmy05})
data set in the 15--150~keV energy band was carried out for the first time in the
GRB rest-frame average, thanks to the large number of GRBs detected by Swift with measured redshift.
No significant differences were found between the observer and the rest-frame behaviour
(\citealt{Guidorzi12}; hereafter, G12).
Notably, no evidence for the break around 1--2~Hz was found in the 15--150~keV band.
In the present work we aim to study the average PDS in two different unexplored regimes with two
different data sets. The goal of this analysis is twofold: i) we address the same average PDS analysis
through two additional data sets from independent satellites and detectors;
ii) these data sets allow us to study the average PDS at very high frequency (up to 1~kHz) with the
{\em BeppoSAX}/Gamma--Ray Burst Monitor (GRBM) and across a broad energy band such that of
{\em Fermi}/Gamma-ray Burst Monitor (GBM) from 8~keV to 1~MeV.

In Section~\ref{sec:data} we report the sample selection criteria and the data analysis procedure.
Results are presented in Section 3, followed by discussion and conclusions respectively in
Sections~\ref{sec:disc} and \ref{sec:concl}.
Uncertainties on best-fitting parameters are given at 90\% confidence for one interesting
parameter unless stated otherwise.

\section{Data analysis}
\label{sec:data}

\subsection{\textit{Fermi}/GBM data selection}
\label{sec:data_sel_fermi}
\begin{figure}
\centering
\includegraphics[width=8.5cm]{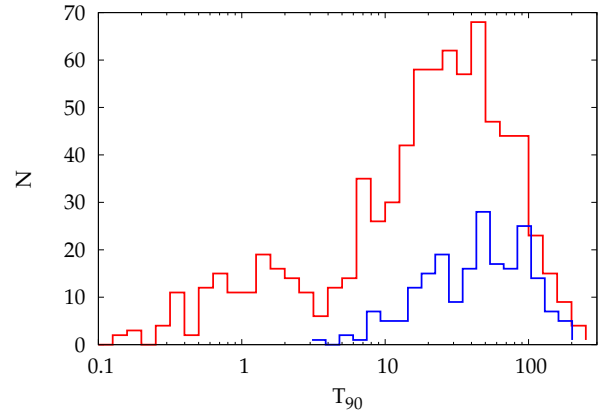}
\caption{$T_{90}$ distributions of a sample of 786 GRBs detected by {\em Fermi}/GBM in the 8--1000~keV energy band
and of the subsample of 205 long GRBs selected for the analysis of the present work. 126 GRBs have $T_{90}<3$~s,
corresponding to $\sim$ 16\% of the whole sample.}
\label{fig01}
\end{figure}

We initially started with 829 GRBs detected and covered by GBM from July 2008 to December 2011.
For each GRB we took the two most illuminated NaI detectors, for which we extracted the corresponding
light curves with 64~ms resolution, which we then added to have a single light curve. 
In this early stage we considered the Time Tagged Event (TTE) files, which hold information about
trigger time and energy channel of each detected photon. We excluded all GRBs with no TTE file.
In some cases the TTE data do not cover the whole event and thus were not considered for the present analysis.
The GRBs durations were expressed in terms of $T_{90}$, which we estimated from the background-subtracted
light curves (Figure~\ref{fig01}). Background subtraction was performed through interpolation using a polynomial
of either first or second order.
%

We excluded short duration bursts by requiring $T_{90}>3$~s. At this stage we were left with a
sample of 650 GRBs. We then rejected all the GRBs with a poor signal--to--noise ratio (S/N) excluding
those with peak rate less than $50$~count~s$^{-1}$. 
Spikes caused by radiative decay of some particles dragged in the Earth magnetic field that interact
with the spacecraft payload were observed in 22 light curves, whose GRBs were therefore rejected
from our sample \citep{Meegan09}.

The extraction of the light curves for each GRB in different energy ranges was made retrieving the data
\footnote{http://legacy.gsfc.nasa.gov/fermi/data/gbm/burst} and processing them with the {\sc heasoft} package
(v6.12) following the {\em Fermi} team threads \footnote{http://fermi.gsfc.nasa.gov/ssc/data/analysis/scitools/gbm\_grb\_analysis.html}.
We selected different energy ranges using the tool {\tt fselect}.
We considered the total energy range of the NaI detectors (8--1000~keV) and three main sub-bands
(8--40, 40--200, 200--1000~keV). Light curves were extracted using the {\tt gtbin} tool.
Finally we calculated the PDS for each GRB of the resulting sample in the time interval
from the earliest to the latest bin whose counts exceed the 5$\sigma$ signal threshold above background
(hereafter, $T_{5\sigma}$). 
Table~\ref{tbl-1} reports the time interval and peak count rate for each selected GRB in the 8--1000~keV band.
Moreover, we also selected a subsample of events with S/N$\ge60$ to better explore the high--frequency
behaviour. For this sample we extracted the light curves with a time resolution of $0.5$ ms
(hereafter, very high resolution or VHR curves) both in the same energy band explored by the GRBM
(40--700 keV) and in the total NaI energy band (8--1000 keV).

\begin{table}
\centering
\caption{Time and Peak count rate. {\em Fermi}/GBM full sample including 205 GRBs.
The PDS is calculated in the time interval reported.
This table is available in its entirety in a machine-readable form
in the online journal. A portion is shown here for guidance.}
\label{tbl-1}
\begin{tabular}{lrrrr}
\hline
Trigger & $t_{\rm start}^{\mathrm{(a)}}$ & $t_{\rm stop}^{\mathrm{(a)}}$ & Peak rate        & $T_{90}$\\
        & (s)                       & (s)                      & (count~s$^{-1}$)  & (s)\\
\hline
080714745 & $ -1.76 $ & $ 31.77 $ & $ 69.3 \pm 4.8 $  &  $ 39 $\\
080723557 & $ -0.13 $ & $ 119.42 $ & $ 460 \pm 17 $  &  $ 77 $\\
080723985 & $ -0.29 $ & $ 52.89 $ & $ 127.6 \pm 8.1 $  &  $ 43 $\\
080724401 & $ -0.11 $ & $ 48.34 $ & $ 268 \pm 20 $  &  $ 42 $\\
080730786 & $ -0.91 $ & $ 18.54 $ & $ 233 \pm 14 $  &  $ 18 $\\
080806896 & $ -6.98 $ & $ 40.63 $ & $ 113.1 \pm 7.9 $  &  $ 44 $\\
080807993 & $ 0.01 $ & $ 49.86 $ & $ 267 \pm 20 $  &  $ 20 $\\
080810549 & $ -10.59 $ & $ 102.31 $ & $ 50.0 \pm 4.5 $  &  $ 58 $\\
080816503 & $ -0.47 $ & $ 69.35 $ & $ 122 \pm 11 $  &  $ 65 $\\
080816989 & $ 0.04 $ & $ 29.10 $ & $ 98 \pm 11 $  &  $ 6 $\\
\hline
\end{tabular}
\begin{list}{}{}
\item[$^{\mathrm{(a)}}$]{Referred to the {\em Fermi}/GBM trigger time.}
\end{list}
\end{table}

We then subtracted the white noise and checked its Poissonian nature related to the statistical
fluctuations observed in light curves. To check the Poissonian character of noise we estimated the mean power
at $f>6$~Hz (Table ~\ref{tbl-2}) and compared it against the value of 2, namely the expected value of a
$\chi^2_2$-distribution for pure Poissonian variance in the Leahy normalisation \citep{Leahy83}.
%
\begin{table}
\centering
\caption{White noise level (Leahy normalisation) for the full {\em Fermi} sample.
The mean value of this sample is $1.99\pm0.02$. Uncertainties at 1$\sigma$.
This table is available in its entirety in a machine-readable form
in the online journal. A portion is shown here for guidance.}
\label{tbl-2}
\begin{tabular}{lc}
\hline
Trigger & $\langle P\rangle$\\
        &  ($f>6$~Hz)\\
\hline
080714745 & $ 1.63 \pm 0.42 $\\
080723557 & $ 2.00 \pm 0.24 $\\
080723985 & $ 2.50 \pm 0.38 $\\
080724401 & $ 2.26 \pm 0.39 $\\
080730786 & $ 2.44 \pm 0.63 $\\
080806896 & $ 1.67 \pm 0.35 $\\
080807993 & $ 2.05 \pm 0.24 $\\
080810549 & $ 1.98 \pm 0.24 $\\
080816503 & $ 2.09 \pm 0.31 $\\
080816989 & $ 2.31 \pm 0.51 $\\
\hline
\end{tabular}
\begin{list}{}{}
\item[$^{\mathrm{(a)}}$]{Too low statistic at $f>6$~Hz. In this case white noise start at lower frequency, so we have estimated the $\langle P\rangle$ level above 4$~Hz$.}
\end{list}
\end{table}
Furthermore, we grouped the background-subtracted PDS along frequency so as to fulfil a 3$\sigma$ significance
criterion for each grouped bin. Following the same procedure by G12 for the {\em Swift}/BAT data,
the selection excluded in each sample (total, low, middle and high energy range) the GRBs whose
grouped PDS collected less than 4 grouped frequency bins.

We ended up with 205 GRBs that will be referred to as the {\em Fermi} sample with a 64~ms time resolution in the total energy range
and, respectively, {we ended up with 155, 201 and 74 in the three energy sub-bands}: 8--40, 40--200, and 200--1000~keV
(low, middle and high energies). The VHR sample includes 96 GRBs whose light curves were extracted in the
8--1000~keV and 40--700~keV energy bands.
For each of these samples we calculated and modelled the average PDS.

\subsection{\textit{BeppoSAX}/GRBM data selection}
\label{sec:data_sel_sax}
For the {\em BeppoSAX}/GRBM GRB sample we started from the GRB catalogue \citep{Frontera09} by selecting
the GRBs fully covered by the high time resolution mode, available only for those which triggered the GRBM
on-board logic. We then excluded the GRBs whose light curves were hampered by gaps in the time profiles.
Finally we selected the GRBs with the highest S/N ($>40$) and ended up with a sample of 89 GRBs.
This requirement was motivated by the need of having very good statistical quality even at high frequencies.

Two different kinds of time resolution are available in the GRBM data: i) light curves with $7.8125$~ms resolution
from $-8$ to $98$~s from the on-board trigger time (hereafter, these curves are referred to as high-resolution
or HR curves); ii) light curves with $\sim0.5$~ms for the first 10~s
from the trigger time (VHR curves). 
Therefore the corresponding Nyquist frequencies are respectively 64~Hz and 1~kHz.
The VHR light curve can be obtained only for a sub-sample of 74 GRBs
For each GRB we extracted the PDS in two different time intervals, depending on the type of light curve:
the PDS of the HR curves was extracted on the $T_{5\sigma}$, like in the case of {\em Fermi}/GBM data
(Section~\ref{sec:data_sel_fermi}), whereas that of the VHR curves was forcibly bound to the first 10~s
from the trigger time.
Table~\ref{tbl-3} reports the time interval and peak count rate for each selected GRB of the HR set.
Also for {\em BeppoSAX} data the final PDS obtained for each GRB of each sub-sample was grouped according
to a 3-$\sigma$
significance criterion excluding the events of the HR sample with fewer than 4 grouped bins and those with
of the VHR sample with fewer than 10 bins. Consequently, the final samples include 42 GRBs with HR data and
25 GRBs with VHR data. Hereafter, the two samples are referred to as the {\em BeppoSAX} HR and the VHR sample,
respectively.

\subsection{PDS calculation}
\label{sec:pds_calc}
Each PDS was calculated through the mixed-radix FFT algorithm implemented within the GNU Scientific Library
\citep{Galassi09},\footnote{http://www.gnu.org/s/gsl/} which does not require the total number of bins to
be a power of 2 \citep{Temperton83} similarly to what was done for the \textit{Swift}/BAT sample (G12).
We calculated the PDS for each GRBs adopting the Leahy normalisation.
For each individual PDS the background level, corresponding to the white noise due to counting statistics, was
initially estimated by fitting with a constant the high-frequency range, where the signal is negligible with
respect to the statistical noise.

Within the Leahy normalisation, a pure Poissonian noise corresponds to a power value of 2.
Therefore we checked the high-frequency constant value for the power averaged out among all the PDSs.
For {\em Fermi} sample the mean value of white noise level is estimated at $1.99\pm0.02$ for $f>6$~Hz,
fully consistent with a Poissonian variance.
For the {\em BeppoSAX} samples the PDS shows evidence for the presence of a small, significant extra-Poissonian
variance of $(3.7\pm1.2)$\% and $(0.94\pm0.35)$\% for the HR and the VHR samples, respectively, in addition to
the statistical white noise. These values were estimated in the frequency range above 50~Hz.

The statistical noise was removed in two different way for different cases. For the {\em Fermi} sample, noise was
assumed to be perfectly Poissonian, compatibly with what we found above. Instead, for the {\em BeppoSAX} samples
it was obtained from fitting the PDS with a constant value estimated at sufficiently high frequencies ($f>50$~Hz)
for each event of the HR sample. 
The estimated background levels are reported in Table ~\ref{tbl-4}. As can be seen in Table ~\ref{tbl-5} for
VHR data, the white noise becomes dominant already at $f>30$~Hz (at higher frequency compared to
the {\em Fermi} case). Indeed, we did not find significantly different
values for the mean power between the two following frequency ranges: $f>30$~Hz and $f>50$~Hz.

After calculating the white noise level for each GRB, we subtracted it and renormalised the PDS by the
corresponding net variance (G12). This choice ensures that all GRBs have equal weights in the average PDS.

The binning scheme used to average the PDS is different for each considered sample.
In the {\em Fermi} case with 64--ms binning time the Nyquist frequency is $7.8125$~Hz,
so we defined a uniform frequency binning scheme with a step of $0.01$~Hz.
At $f<0.01$~Hz we considered two bins, $0.001$~Hz~$\le f<0.005$~Hz and
$0.005$~Hz $\le f<0.01$~Hz. The same step is used in the frequency grid defined for the average PDS
of the HR {\em BeppoSAX} data. In the {\em BeppoSAX} case the PDS have correspondingly more frequency bins,
due to the higher Nyquist frequency. We took only one single bin from $0.001$~Hz and $0.01$~Hz.
The frequency grid changes for the VHR data: we chose a broader frequency step of $1$~Hz because the total
PDS extraction time is limited to 10~s for each {\em BeppoSAX} light curve and this implies a coarser
frequency resolution. For the VHR PDS we considered 4 bins with step of $0.2$~Hz at $f<1$~Hz.
For each individual GRB we calculated the average power in each frequency bin of the corresponding grid
described above. Finally, for each frequency bin of the grid we determined the average power over all GRBs
of a given sample after they had been renormalised.
Finally the frequency bins of the average noise--subtracted PDS were grouped by requiring at least 3$\sigma$
significance to reduce the uncertainties at high frequencies.

\subsection{PDS fitting}
\label{sec:pds_fit}
The average PDS was modelled using a smoothly broken power-law in the same parametrisation as that adopted
by G12,
\begin{equation}
{\rm PDS}(f) = 2^{1/n}\,F_0\ \Big[ \Big(\frac{f}{f_{\rm b}}\Big)^{n\,\alpha_1} +
\Big(\frac{f}{f_{\rm b}}\Big)^{n\,\alpha_2} \Big]^{-1/n}
\qquad ,
\label{eq:mod}
\end{equation}
where the parameters left free to vary are the break frequency $f_{\rm b}$, the two power-law indices $\alpha_1$
and $\alpha_2$ ($\alpha_2>\alpha_1$) and the normalisation parameter, $F_0$. The smoothness parameter $n$ could not
be effectively constrained in all cases, thus it was fixed to $n=10$, corresponding to a relatively sharp break
around $f_{\rm b}$, for all cases to ensure a more homogeneous comparison between the best-fit values obtained
over different sets as well as with previous results obtained from the {\em Swift} data.
Thanks to the central limit theorem, we can assume these variables to be normally distributed. This allowed
us to determine the best-fitting model by minimising the following un-normalised negative log--likelihood
function,
\begin{equation}
\centering
L = \frac{1}{2}\ \sum_{i=1}^{N_f}\Big(\frac{P_i-{\rm PDS}(f_i)}{\sigma_i^2}\Big)^2
\qquad ,
\label{eq:likel}
\end{equation}
where $P_i$ and $f_i$ are the observed power and frequency of the $i$-th bin.
$N_f$ is the number of frequency bins, excluding the Nyquist frequency.

%
\section{Results}
\label{sec:res}
\subsection{Average PDS at different energy bands}
\label{sec:res_fermi}

Table~\ref{tbl-6} reports the best-fit parameters estimated for the average PDSs for the different GRB
samples considered.

\begin{figure}
\centering
\includegraphics[width=8.5cm]{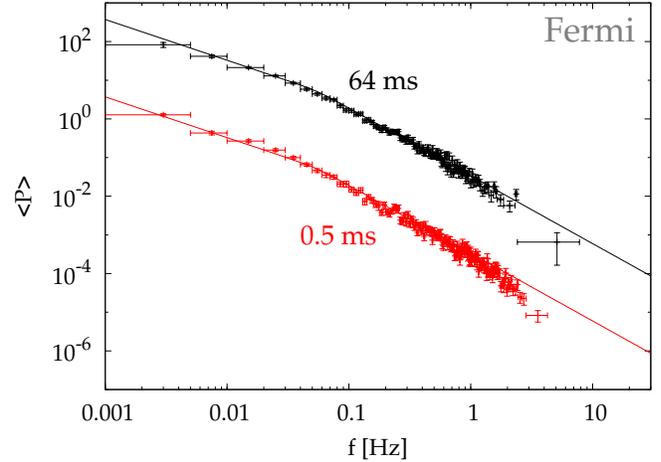}
\caption{Top (bottom) data show the average PDS for a sample of 205 (96) {\em Fermi}/GBM GRBs
in the 8--1000~keV energy range with  64--ms ($0.5$--ms) time resolution.
Solid lines show the best--fitting model obtained on the 64--ms data set and the same model
renormalised to the $0.5$--ms data set, which was shifted for the sake of clarity.
The bottom data set clearly deviates from the model at $f\ga1$~Hz.}
\label{fig02}
\end{figure}

For the average {\em Fermi} PDS extracted in the total energy range 8--1000~keV (Figure~\ref{fig02}) with 64--ms binning time
the best-fitting parameters are $\alpha_1=1.06_{-0.07}^{+0.05}$, a break at $5.5\times10^{-2}$~Hz above which
the PDS steepens to $\alpha_2=1.75\pm0.03$. This slope of the spectra is very similar to the previous values
found in the literature related to the GRBs detected with BATSE in similar energy bands (BSS98, BSS00),
and in agreement with the value of $5/3$ of a Kolmogorov spectrum.
%
%
%
Indeed BSS00 have found an index ranging from $1.50$ to $1.72$ in the frequency range $0.025<f<1$~Hz
fitting the average PDS resulted from the BATSE sample (20--2000~keV) with a simple power--law.
Moreover, also for the average PDS of {\em Swift}/BAT data (15--150 keV) we see a typical slope
described with a low--frequency index $\alpha_1=1.03\pm0.05$ up to a break frequency around
$3\times10^{-2}$~Hz, followed by and an index $\alpha_2=1.73_{-0.03}^{+0.04}$ (G12).
Since the break frequency $f_{\rm b}$ is sensitive to the average characteristic time $\tau$ of
typical individual shots roughly as $f_{\rm b}\sim 1/(2\pi\tau)$ \citep{Frontera79,Belli92,Lazzati02},
the value we found in the {\em Fermi} data corresponds to a mean characteristic time of about 3~s.

Comparing the average PDS of the whole Fermi sample with that of the high--quality (S/N$\ge60$)
subsample extracted with $0.5$--ms resolution, the latter data set shows evidence for a further break
around 1--2~Hz with respect to the best--fitting model obtained for the former data set (bottom data
in Fig.~\ref{fig02}). The behaviour of the average PDS at high frequency is thoroughly
discussed in Section~\ref{sec:res_sax} together with {\em BeppoSAX} data.

\begin{figure}
\centering
\includegraphics[width=8.5cm]{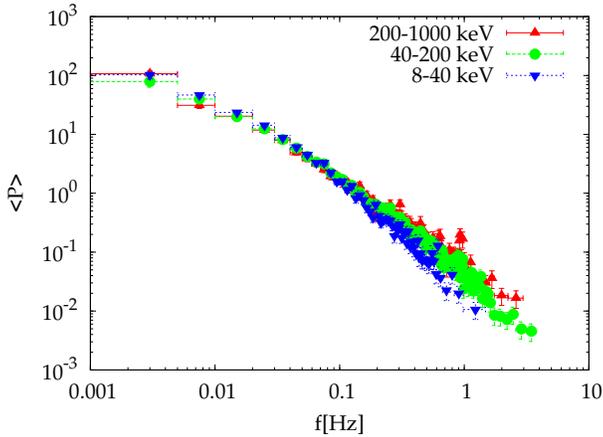}
\caption{Different slopes observed in the average PDS extracted on the three main energy ranges using a time resolution of 64~ms. Upside--down triangles, circles, and triangles show 
the 8--40, 40--200, and 200--1000~keV energy bands, respectively.
The spectrum becomes shallower moving from low to high energies.}
\label{fig03}
\end{figure}

The analysis of the average PDS at different energy channels reveals a clear trend of the spectral shape
when we move from soft to hard energy ranges. Figure~\ref{fig03} displays the average PDS corresponding
to three different energy channels: 8--40, 40--200, and 200--1000~keV.
The index $\alpha_2$ decreases from $1.95$ to $1.47$ moving from 8--40 to 200--1000~keV.
This reflects the known narrowing of pulses with energy, according to which the same GRB pulse appears
to be narrower and spikier at higher energies \citep{Fenimore95,Norris96,Piro98}.
The same trend was observed in the BATSE average PDS (BSS00), for which the power--law index decreases
from $1.72$ in the 25--55~keV to $1.50$ above 320~keV. Furthermore, a similar behaviour is observed in
the {\em Swift} data, with $\alpha_2$ varying from $1.75_{-0.04}^{+0.05}$ to $1.49_{-0.07}^{+0.08}$ passing from
15--50 to 50--150~keV.

\begin{figure}
\centering
\includegraphics[width=8.5cm]{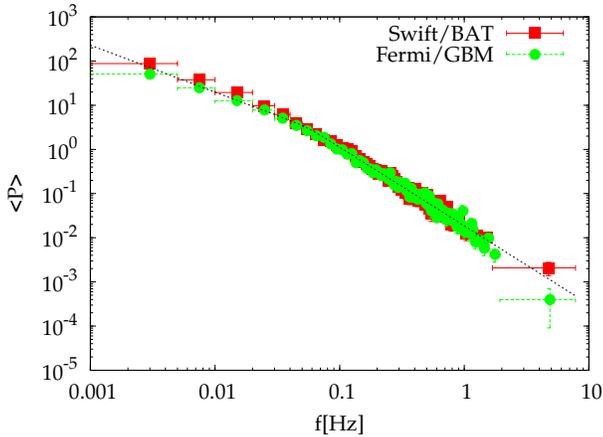}
\caption{Average PDS of {\em Fermi} sample in the 15--150~keV energy range (circles) compared with the
{\em Swift}/BAT result (squares) provided by G12. Both are calculated from 64--ms binned light curves.
The two independent measures are compatible. The dashed line shows the best fit model for Fermi data.}
\label{fig04}
\end{figure}

We also extracted the light curves in the common energy bands with other instruments so that we can compare
results limiting the systematic differences connected with different energy passbands.
The average {\em Fermi}/GBM PDS obtained in the typical {\em Swift}/BAT energy range (15--150~keV) are perfectly consistent with
the analogous results on {\em Swift}/BAT data, as shown in Fig.~\ref{fig04}.
The slope indices of average {\em Fermi}/GBM PDS for the 15--150~keV are $\alpha_1=1.06_{-0.07}^{+0.06}$ and
$\alpha_2=1.78_{-0.03}^{+0.04}$, to be compared with their analogous values found with {\em Swift}/BAT,
$\alpha_1=1.03\pm0.05$, $\alpha_2=1.73\pm0.03$. So the apparently different values at low frequencies
between the two spectra in Fig.~\ref{fig04} is not statistically significant.

\subsection{FRED sub-sample}
\label{sec:res_fermi_fred}
We investigated whether the GRBs whose light curves can be described as a single
fast rise exponential decay (FRED) show distinctive features in the average PDS.
To this aim, we selected 10~GRB of this kind out of the {\em Fermi} sample
by visual inspection and calculated the corresponding average PDS.
The best-fit parameters in this case are $\alpha_1=1.32\pm0.10$ and $\alpha_2=2.53_{-0.24}^{+0.39}$
with a break at about $6\times10^{-2}$~Hz (see Table~\ref{tbl-6}).
That the high-frequency tail of the PDS for
the FRED sample is steeper than that of the whole sample of GRBs, agrees with
the PDS expected for a single FRED (e.g., see Lazzati 2002).\nocite{Lazzati02}
This in turns suggests that the average PDS of multiple--pulse GRBs is shallower
because of the presence of various characteristic times. The sum of several PDS
with different break frequencies would therefore result in a simple power--law
with no dominant break in the explored frequency range.

\begin{figure}
\centering
\includegraphics[scale=0.3]{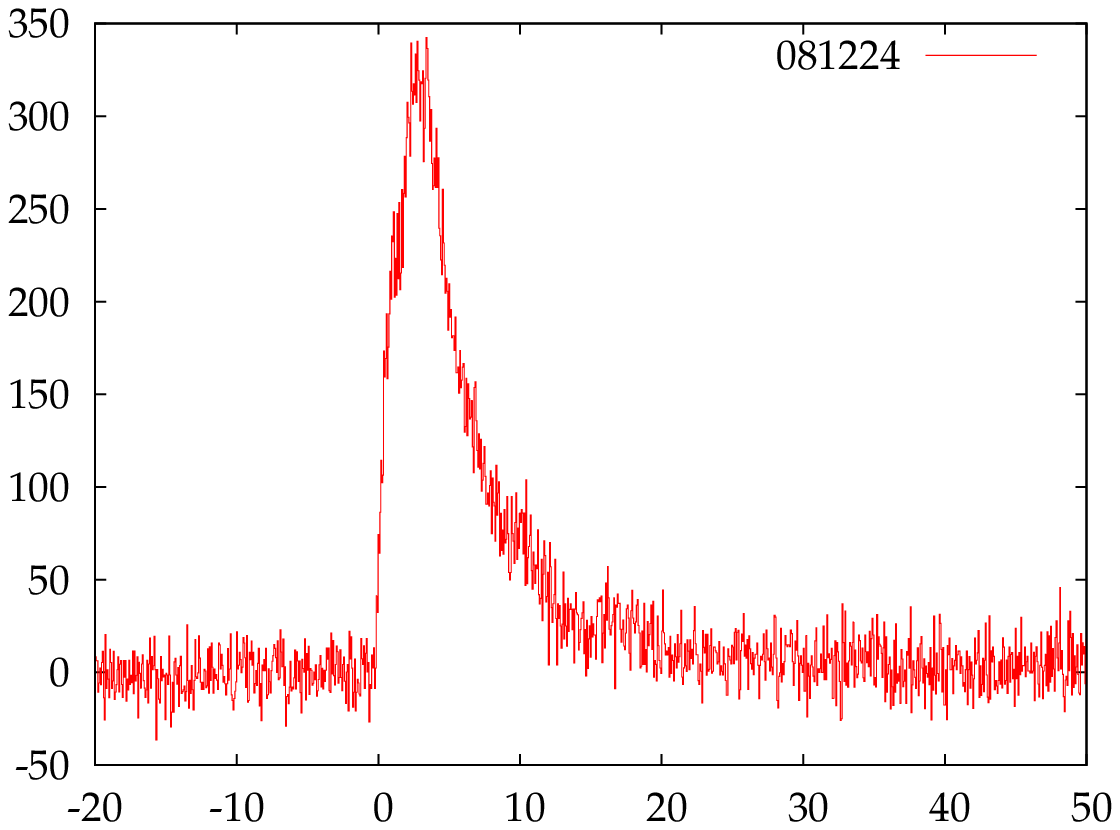}
\includegraphics[scale=0.3]{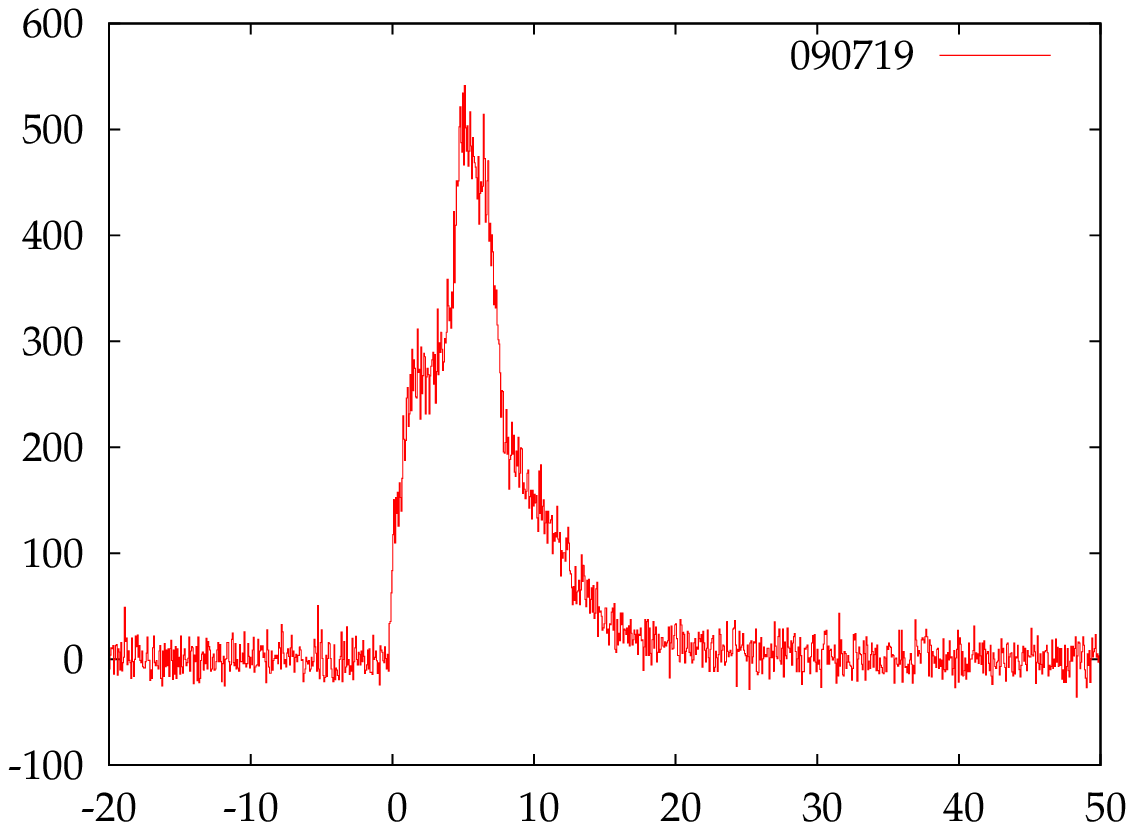}
\includegraphics[scale=0.3]{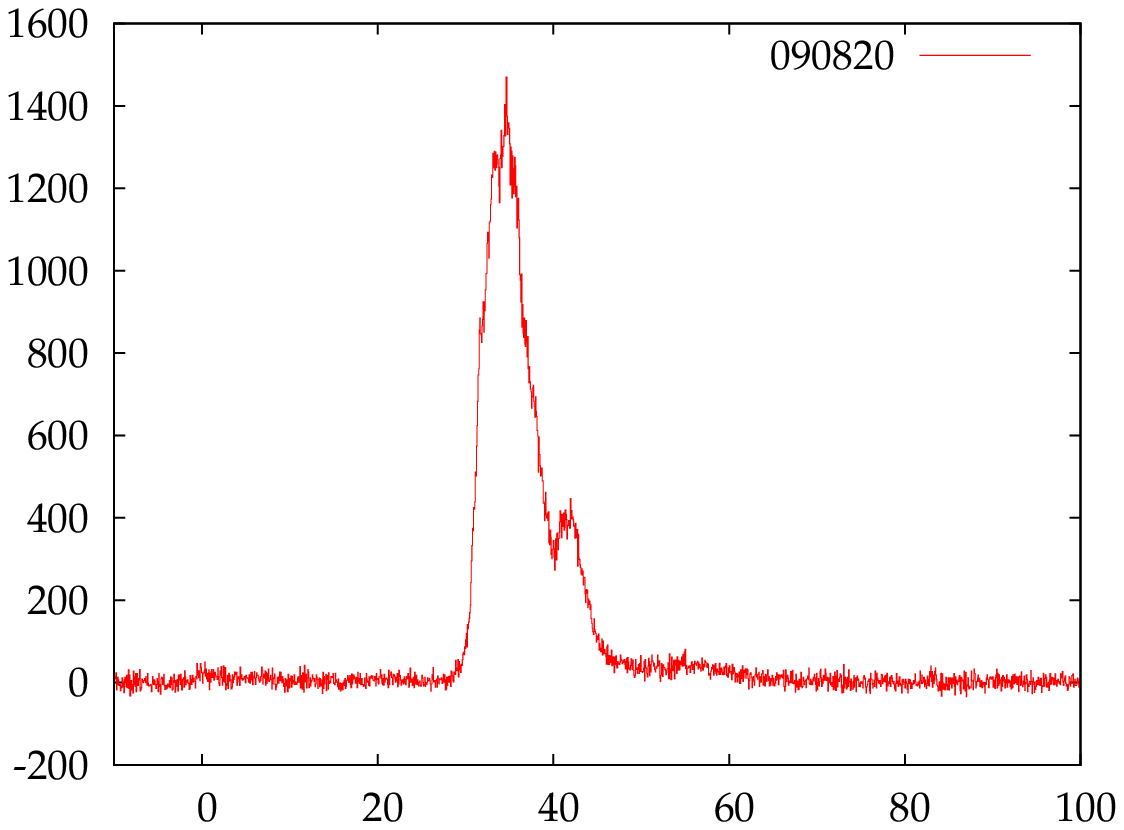}
\includegraphics[scale=0.3]{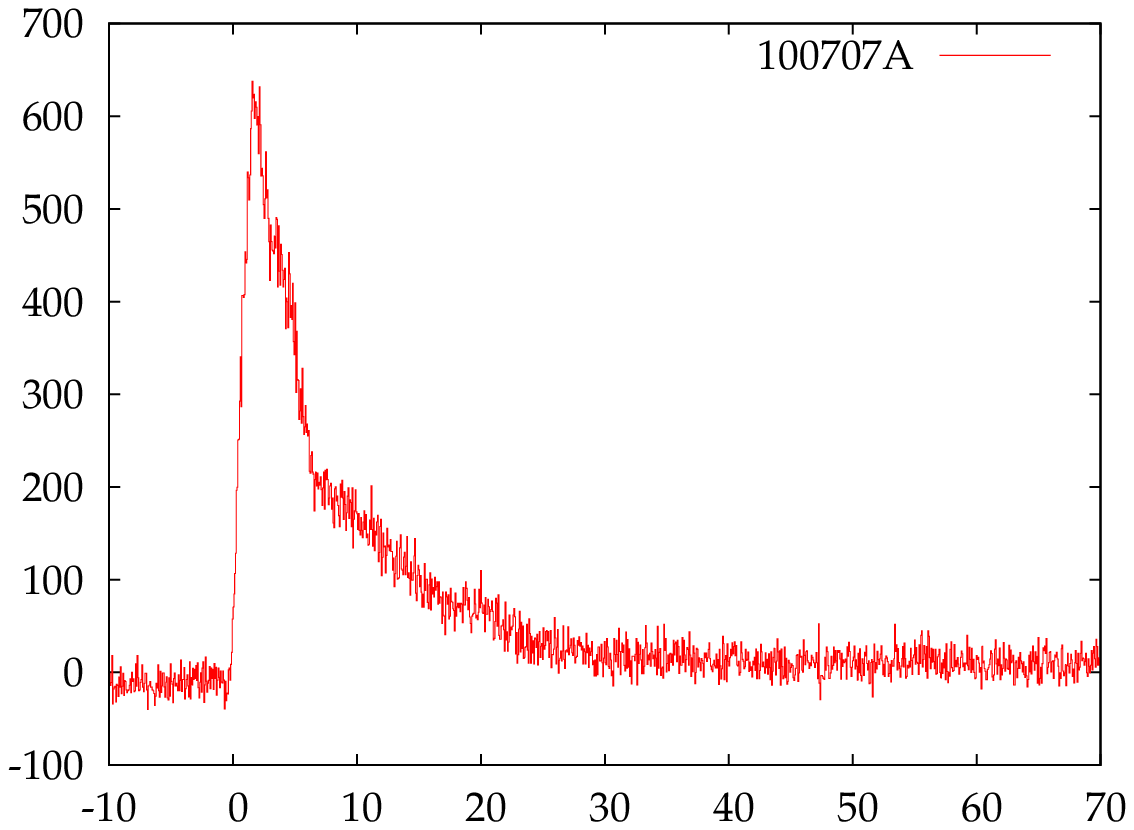}
\includegraphics[scale=0.3]{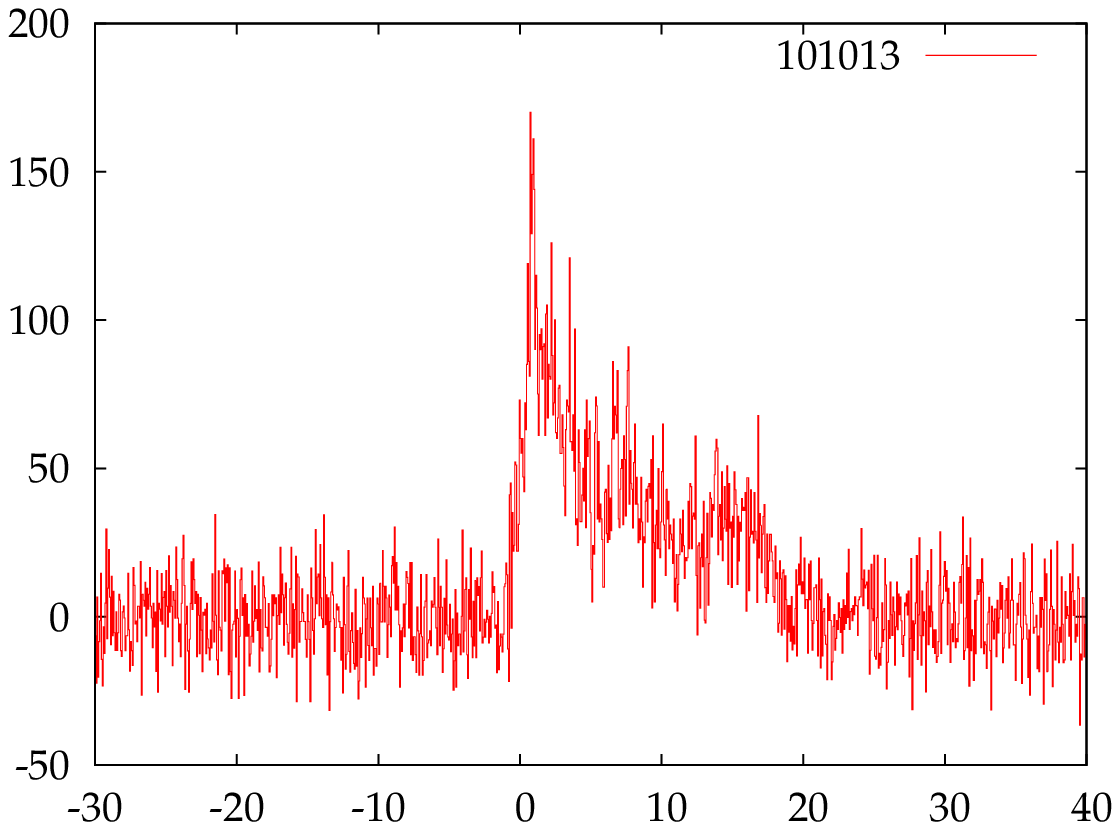}
\includegraphics[scale=0.3]{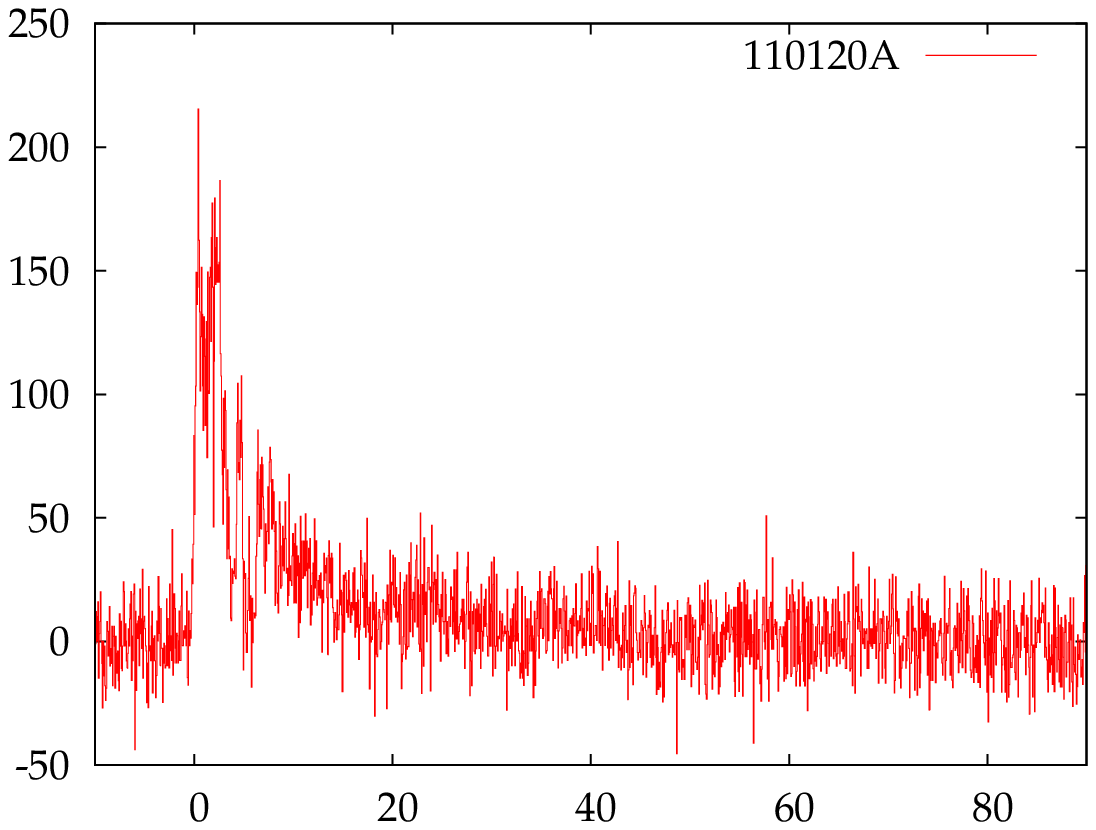}
\includegraphics[scale=0.3]{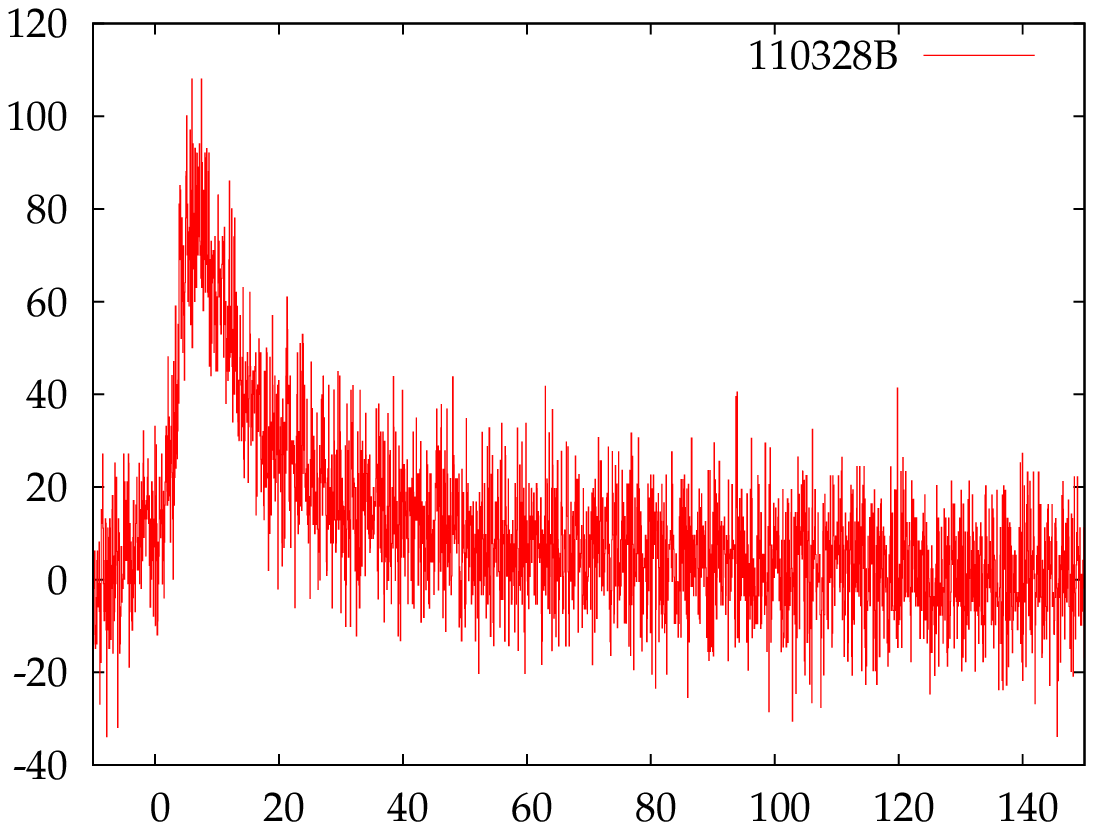}
\includegraphics[scale=0.3]{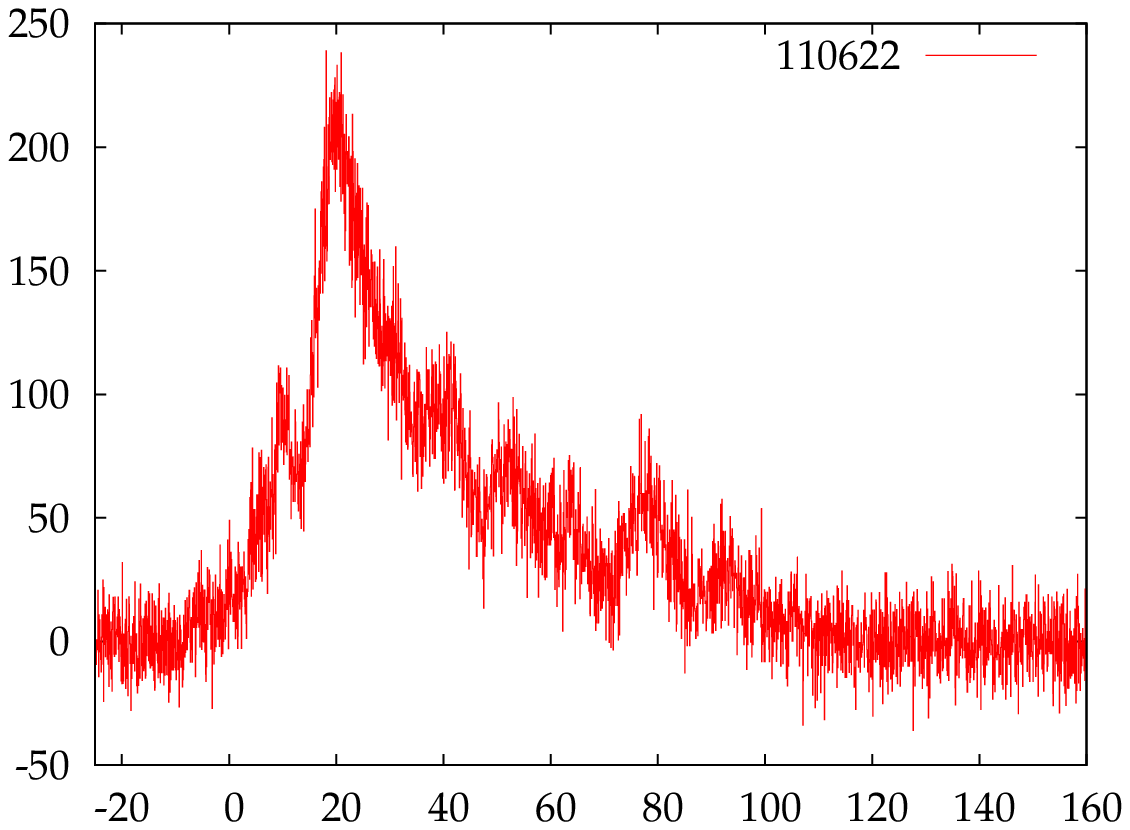}
\includegraphics[scale=0.3]{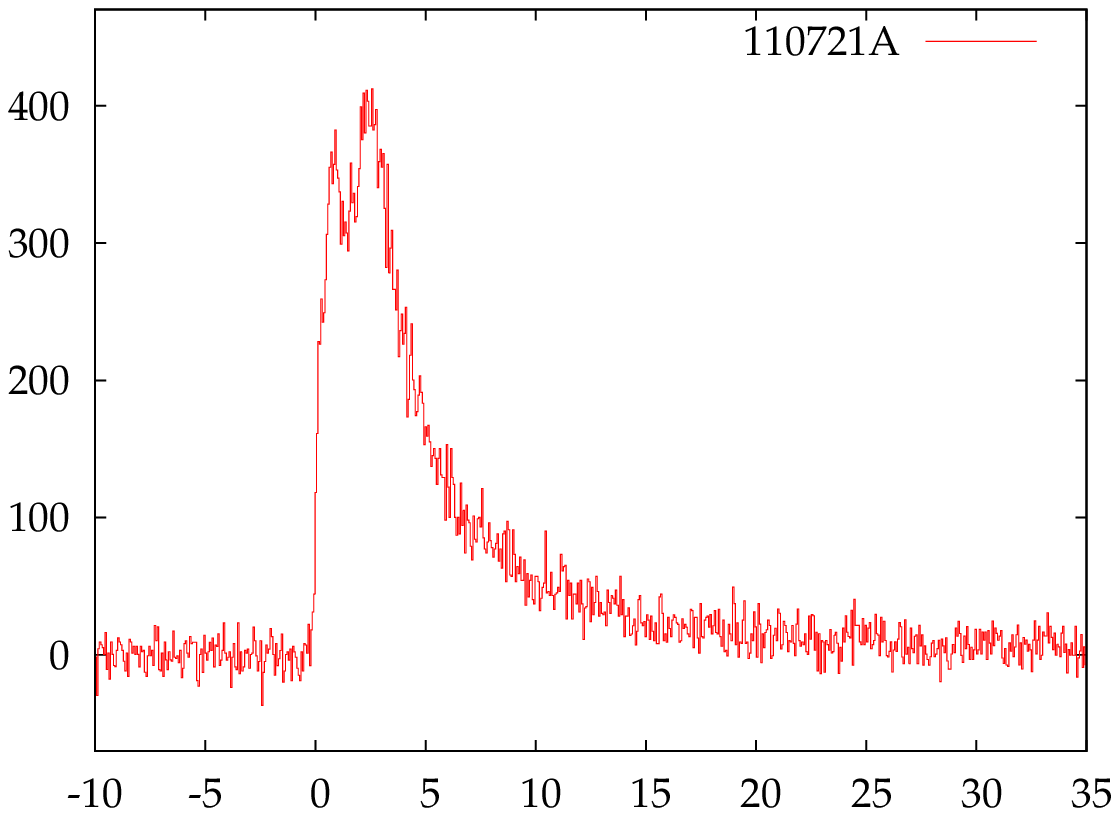}
\includegraphics[scale=0.3]{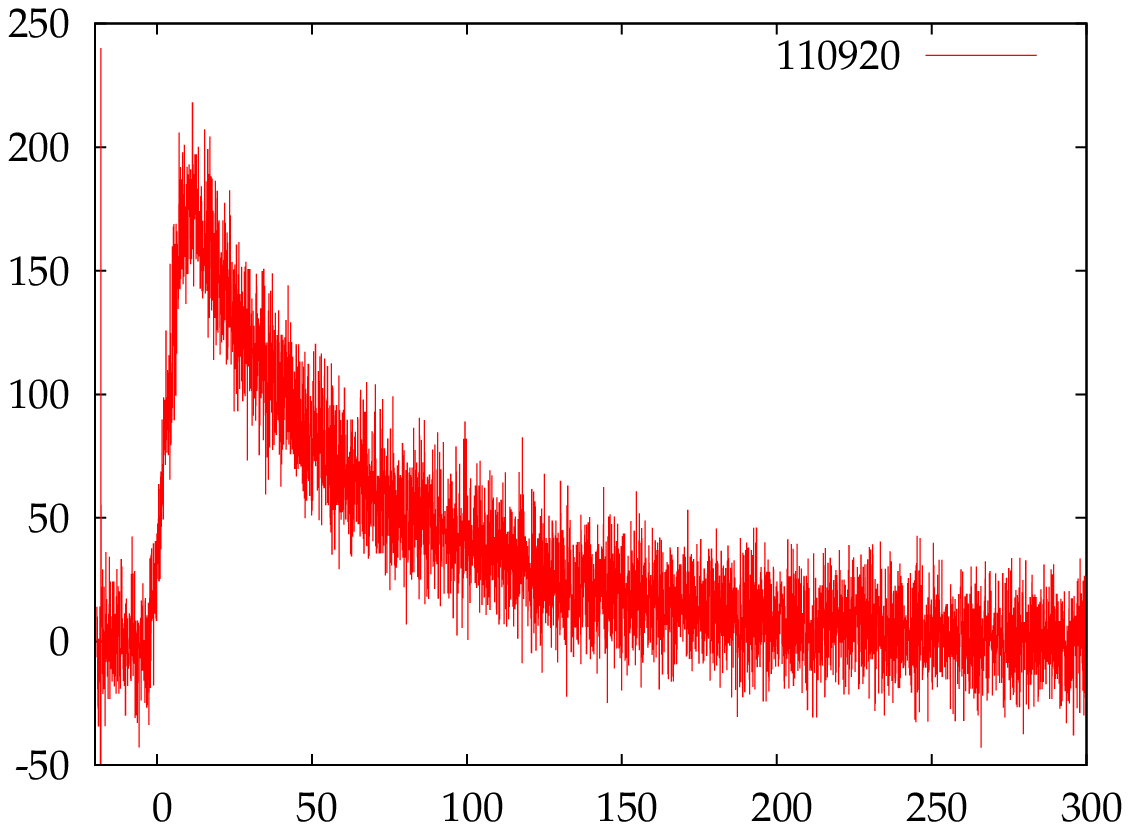}
\caption{The 10 FRED used in our PDS analysis. Each curve has a 64--ms bin time and is expressed in
units of counts~s$^{-1}$ as a function of time.}
\label{fig05}
\end{figure}

\begin{figure}
\centering
\includegraphics[width=8.5cm]{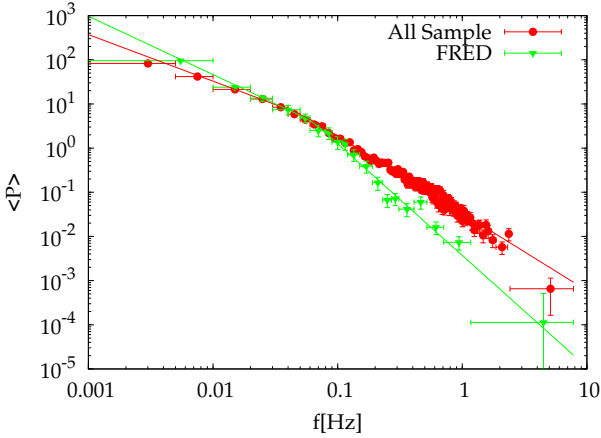}
\caption{The average PDS of the FRED sample (detected with GBM) is shallower than the
average PDS of the full {\em Fermi} sample. The energy band is $8$--$1000$~keV for
both sets with 64~ms time resolution.}
\label{fig06}
\end{figure}

\subsection{Average PDS up to high frequency}
\label{sec:res_sax}

The average PDS for the HR light curves provided by the {\em BeppoSAX}/GRBM shows a second
break at high frequency ($f_{\rm b2}=1.91_{-0.43}^{+0.45}$~Hz). The observed slope can be parametrised
with two indices, $\alpha_2=1.49\pm0.04$ and $\alpha_3=2.46_{-0.31}^{+0.44}$ (we did not use $\alpha_1$,
which has previously been used to denote the slope below a few $10^{-2}$~Hz).
The break is likely to be real because the improvement is significant compared to
the same model without it. The F--test yields a probability of $1.26$\% that the break is not required.
The value itself of this break frequency as well as the values of the corresponding power--law indices
indicate that this feature has a different origin from the other one observed at lower frequency.
This motivated us to adopt a different notation for the power-law index above this break, $\alpha_3$.
Overall, the different slopes refer to the corresponding frequency ranges: $\alpha_1$ below a few
$10^{-2}$~Hz, $\alpha_2$ holds in the range $10^{-2}<f<1$~Hz, and $\alpha_3$ for $f\ga1$~Hz.
 
\begin{figure}
\centering
\includegraphics[width=8.5cm]{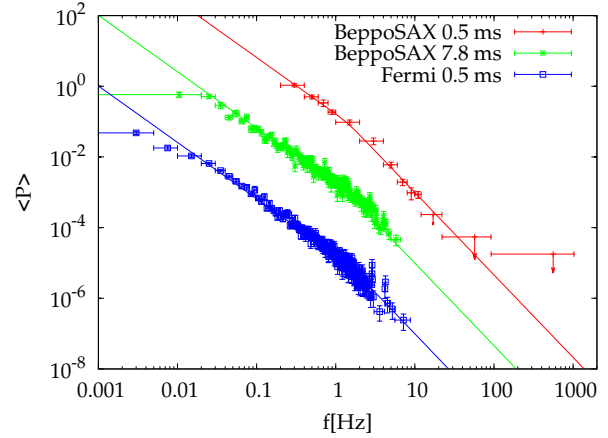}
\caption{The top, mid, and bottom average PDS refer to the
{\em BeppoSAX} $0.5$--ms, $7.8$--ms resolution samples and to the
{\em Fermi} subsample with S/N$\ge60$, respectively, for the 40--700~keV band.
The break around 1--2~Hz is present in each set.
The common best--fitting model is also shown, for both the {\em BeppoSAX} and
{\em Fermi} data sets. The corresponding model parameters were bound to be equal
for all the data sets, except for the normalisation terms.
Upper limits are given at 2$\sigma$ confidence. Data were shifted vertically
for the sake of clarity.}
\label{fig07}
\end{figure}

When we limit our PDS analysis to the first 10~s of the GRBM trigger time of each GRB light curve and use
the VHR data, a very similar result is found for the average PDS, which now extends up to 1~kHz.
The best--fitting parameters for these data are $\alpha_2=1.52\pm0.17$, $\alpha_3=2.91_{-0.41}^{+0.51}$ with a clear
break at $f_{\rm b2}=2.59_{-0.94}^{+1.04}$~Hz (Fig.~\ref{fig07}). Also in this case a break in the model
is required to fit the data, with a probability of $0.47$\% that the improvement obtained with the break
is due to chance according to the F--test.
Furthermore in Fig.~\ref{fig02} the average PDS obtained from the {\em Fermi} VHR sample looks like it
also requires a break at $f\ga1$~Hz. To check the mutual compatibility of these data with a broken power--law
model, we extracted the {\em Fermi} VHR average PDS over the same energy range covered also by GRBM,
40--700~keV. To fit these data above $0.02$ Hz we used a simple power--law as well as a broken power--law
and used the F--test to evaluate the improvement one obtains moving from the former to the latter.
We estimate a probability of $3.4$\% that such improvement is just by chance. We found two
different slopes, $\alpha_2=1.65\pm0.03$ and $\alpha_3=2.41_{-0.19}^{+0.34}$, with a break at
$f_{\rm b2}=1.1_{-0.2}^{+0.3}$~Hz ($\chi^2/dof=1.07$).
We excluded from the fit the lowest frequency point in the {\em BeppoSAX} HR PDS and in the
{\em Fermi} VHR PDS (40--700~keV), because it clearly lies below the extrapolation of a double
broken power--law, since it is clearly affected by the presence of the low--frequency break.

We also performed a combined analysis of the two and three samples, {\em BeppoSAX} (HR + VHR)
(i.e., {\em BeppoSAX} data alone), and {\em BeppoSAX} (HR + VHR) plus {\em Fermi} VHR,
fitting all the spectra simultaneously with the same model,
apart from allowing each set a different normalisation term.
For the {\em BeppoSAX} data alone, the resulting break frequency is found to be
$f_{\rm b2}=2.11_{-0.33}^{+0.42}$~Hz, while the two slopes
have indices respectively $\alpha_2=1.50_{-0.04}^{+0.03}$ and $\alpha_3=2.69_{-0.20}^{+0.27}$.
This treatment implicitly assumed the two data sets to be statistically independent. Although this
is not completely true, since the 10~s data of the VHR curves are part of the full profile of about
100~s of HR data, on average the common data amount to 10-20\% or so. Consequently, the expected
correlation between the two data set affects the results within a comparable fraction.
By adding the VHR sample extracted with {\em Fermi}, we found $\alpha_2=1.60_{-0.03}^{+0.02}$,
$\alpha_3=2.33_{-0.13}^{+0.15}$ with a break at $f_{\rm b2}=1.4\pm0.3$~Hz
($\chi^2/{\rm dof}=1.37$).
We tried to see whether the quality of the fit could be improved by allowing the smoothness
parameter to vary (eq.~\ref{eq:mod}), thus allowing a smooth transition from one power--law regime
to the following one, with no appreciable result though.

\begin{figure}
\centering
\includegraphics[width=8.5cm]{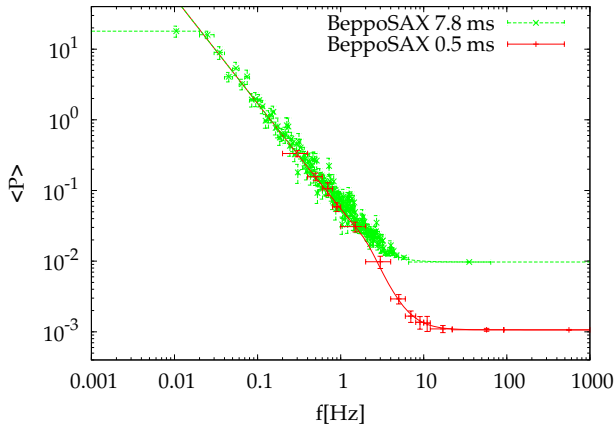}
\caption{The average PDS obtained from {\em BeppoSAX} samples without the white noise subtraction.
The break at 1--2~Hz is still evident thanks to the signal being more than one order of magnitude
higher than the white noise level. This rules out any bias due to possibly wrong white noise subtraction.}
\label{fig08}
\end{figure}

Although the white noise subtraction was done through a careful estimation of the high frequency
power (Section~\ref{sec:pds_calc}), we examined whether the break could be an artifact of a small
bias in the white noise subtraction. More specifically, overestimating the white noise could mimic
the appearance of an artificial break. To test this possibility, we extracted the
average PDS {\it without} noise subtraction, keeping the same relative normalisation for each GRB
as that of the noise--subtracted case. We fixed the best-fitting model of the noise-subtracted
PDS obtained above and fitted the white noise with a constant.
Figure~\ref{fig08} clearly shows that the break in the average PDS occurs when the average signal
still dominates the white noise level (by more than one order of magnitude in the VHR data).
This rules out the possibility of the break around 1--2~Hz being the result of biased white noise
subtraction and suggests it to be a genuine feature of the average PDS at energies above
$40$~keV.

\section{Discussion}
\label{sec:disc}
In general, two distinct sources of time variability have been found to characterise the
GRB variability:
a fast component dominated by the presence of relatively short ($<$~1~s) pulses and
a slow component linked to pulses lasting several seconds \citep{Scargle98,Vetere06,Margutti09,Gao12}.
These two kinds of dominant time scales should be produced by different
mechanisms involved in the physical process, and different explanations in different
scenarios are available in the literature \citep{Morsony10,ICMART,Titarchuk12}.
The simple power--law modelling the average PDS and encompassing nearly two orders of magnitude
in frequency is suggestive of some kind of scale invariance within the same frequency range,
thus confirming the coexistence of multiple characteristic timescales.

The study of the average PDS in different energy ranges made possible by {\em Fermi}/GBM provides
clues to better characterise the different aspects of GRB time variability.
The observed energy dependence of the power--law index of the average PDS, $\alpha_2$, in the frequency
range $10^{-2}<f<1$~Hz confirms and extends the results found with previous work and data sets.
Indeed, in the 8--1000 keV band the average PDS of long GRBs detected with GBM show a broken
power--law behaviour ($\alpha_1=1.06_{-0.07}^{+0.05}$, $\alpha_2=1.73_{-0.03}^{+0.04}$ and
$f_{\rm b}=5.5\times10^{-2}$~Hz) with $\alpha_2$ very close to the slope of average PDS observed
in the BATSE analysis ($\alpha\approx1.67$). 

\begin{figure}
\centering
\includegraphics[width=8.5cm]{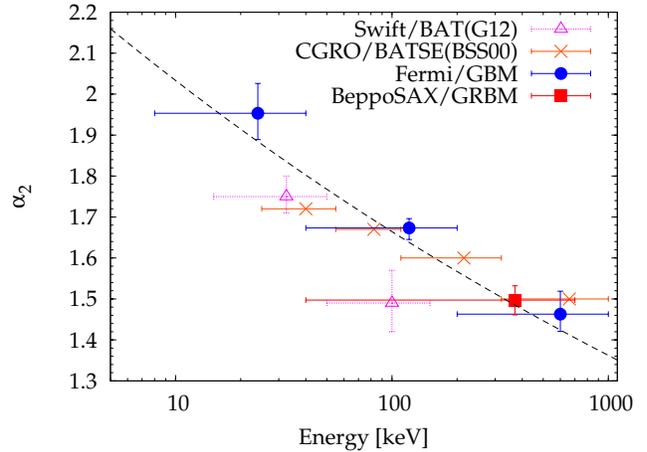}
\caption{The power--law index of the average PDS in the frequency range $10^{-2}<f/{\rm Hz}<1$ obtained
from different data sets as a function of energy. Dashed line ($\alpha_2\propto E^{-0.09}$)
illustrates the $\alpha_2$ dependence on energy as estimated from Fermi data.}
\label{fig09}
\end{figure}

More specifically, the average PDS slope undergoes a steep--to--shallow evolution passing from
soft to hard energy channels, as shown in Fig.~\ref{fig09}.
This behaviour is consistent with the narrowing of pulses with energy: \citet{Fenimore95} found
a dependence of the average pulse width $w$ on energy $E$ as $w\propto E^{-0.4}$, estimated by
measuring the average auto-correlation function (ACF) width for a sample of BATSE bursts as a function
of the energy channel. In addition to the energy dependence of the average pulse width, also
the shape itself and, in particular, the peakedness of the average ACF depends on energy (BSS00).
Indeed, the energy dependence of the shape of the pulse profile explains the energy dependence of the
power--law index:
if the shapes of a given pulse at different energies were the same, only the break frequency in the
average PDS should change correspondingly, while the slope should remain unaffected. Since this is
not what is observed, the evolution with energy of the average power--law index in the PDS confirms
the change in the shape itself of the energy pulse as a function of energy.

Another important result that emerged from the present analysis is the break revealed around 1--2~Hz
in the {\em BeppoSAX} average PDS. Although the evidence for it in the {\em Fermi} data alone
is less compelling because of the lower S/N in that frequency range, the {\em Fermi} average
PDS is fully compatible with it. The joint {\em BeppoSAX}--{\em Fermi} analysis of such high--frequency
break shows that this may significantly vary between 1 and 2~Hz, depending on the GRB sample
and on its average S/N.
Together with results obtained on {\em Swift} data by G12, this break becomes evident at harder energies.

This feature in the average PDS and its possible dependence on energy provides an important clue
to constraining theoretical models proposed to explain the physical mechanism involved in GRBs
and confirms and strengthens the analogous result obtained by BSS00 on BATSE data.
The break could be related to an average intrinsic variability time scale, $\Delta~t\la 0.1$~s,
below which the temporal power changes regime. This may link directly to the central engine.
Alternatively, it could be related to the variation of the outflow Lorentz factor,
or it could depend on the radius at which the expanding shell becomes optically thin $R_*$.
In this latter scenario we could observe variability only on
time scales longer than a characteristic time $t_*=R_*/c\Gamma^2$ (BSS00).

A number of theoretical interpretations of the power--law PDS with an index compatible with
$5/3$ have been put forward in the literature. This is what is expected for a Kolmogorov spectrum
within a medium with fully developed turbulence. For instance, in the internal shock model, the
parameters of the wind of relativistic shells can be constrained so as to reproduce the observed
average PDS \citep{Panaitescu99,Spada00}; or in the context of a relativistic jet making
its way out through the stellar envelope of the progenitor star \citep{Zhang09,Morsony10}.
Within other scenarios, in which the dissipation into gamma--rays is magnetically driven,
the observed features of the average PDS in the frequency range from a few $0.01$ to 1--2~Hz
can also be explained (e.g., Zhang \& Yan 2011).\nocite{ICMART}
Again, alternatively the observed temporal properties could be driven by instabilities in
the accretion disc of potentially different origins: erratic episodic accretion
(e.g., \citealt{Kumar08}); hydrodynamical or magnetic origin
(e.g., \citealt{Perna06,Proga06,Margutti11}).
magneto--rotational origin, in which neutrino cooling is the dominant process \citep{Carballido11}.
The reader is referred to G12 for a more detailed discussion of the predictions of
the various theoretical models with reference to the average PDS properties.

\section{Conclusions}
\label{sec:concl}

We studied the properties of the average PDS of GRBs in two unexplored regimes:
across a broadband energy range from 8~keV to 1~MeV using {\em Fermi}/GBM data
and up to very high frequencies (up to 1~kHz) using {\em BeppoSAX}/GRBM data.

In agreement with previous results obtained from an analogous analysis of {\em CGRO}/BATSE and
of {\em Swift}/BAT data, we also found a clear relation between the average PDS slope from
$\sim0.01$ to $\sim1$~Hz range and energy, with the index spanning the range from $1.5$ to $1.9$
from $8$~keV through $1$~MeV in three channels (8--40, 40--200, and 200--1000~keV).
The slope of the average PDS carries information about the spikiness of light curve as well
as the multiple presence of several characteristic time scales (scale invariance within the two decades
of the aforementioned frequency range).

For the first time we extended the study of PDS up to 1~kHz in frequency with the very high
time resolution provided by {\em BeppoSAX}/GRBM. In this case, the average PDS pinned down a
clear break at 1--2~Hz. This provides a strong clue to the dominant minimum variability
time, potentially connected with either the intrinsic inner engine variability, or with the
dispersion of the bulk Lorentz factor distribution for a wind of relativistic shells, or
with the average distance at which internal collisions dissipate energy into gamma--rays.
Combining our results with those obtained from the {\em Swift} data set, the presence of
this break emerges only in the harder energy channels ($\ga 100$~keV).

The average slope is broadly consistent with the theoretically appealing value of $5/3$ expected
for a Kolmogorov spectrum of velocities within a fully turbulent medium, as suggested in previous works
(BSS98, BSS00).
Our results in the frequency range $\sim10^{-2}$ to $\sim1$~Hz are in broad agreement
with a number of theoretical interpretations within different alternative contexts, encompassing
the classical internal shock scenario as well as the magnetically--dominated outflows models.
Instead, still missing is a detailed theoretical explanation for the other two properties:
i) the presence of the 1--2~Hz break and its energy dependence; ii) the energy dependence of the
average power--law index.

\section*{Acknowledgments}
This work was supported by PRIN MIUR project on ``Gamma Ray Bursts: from progenitors to physics
of the prompt emission process'', P.~I. F. Frontera (Prot. 2009 ERC3HT).
We acknowledge useful comments by the anonymous referee which helped to improve the paper.


\onecolumn
\begin{deluxetable}{lrrrr}
\tablecolumns{5}
\tabletypesize{\scriptsize}
\tablecaption{Time and Peak count rate. {\em BeppoSAX}/GRBM HR sample including 42 GRBs.
\label{tbl-3}}
\tablewidth{0pt}
\tablehead{
\colhead{GRB} & \colhead{$t_{\rm start}$\tablenotemark{a}} & \colhead{$t_{\rm stop}$\tablenotemark{a}} & \colhead{Peak rate} & \colhead{$T_{90}$}\\
 & \colhead{(s)} & \colhead{(s)} & \colhead{(count~s$^{-1}$)} & \colhead{(s)} }
\startdata
970111 & $ -0.34 $ & $ 40.02 $ & $ 46.53 \pm 1.26 $  &  $ 31.00 $\\
970117B & $ -0.25 $ & $ 19.00 $ & $ 61.63 \pm 1.38 $  &  $ 13.00 $\\
970315A & $ -0.41 $ & $ 20.97 $ & $ 116.87 \pm 8.15 $  &  $ 15.00 $\\
970517B & $ -0.75 $ & $ 3.61 $ & $ 139.11 \pm 5.09 $  &  $ 5.00 $\\
970601 & $ 6.99 $ & $ 41.75 $ & $ 74.83 \pm 3.13 $  &  $ 30.00 $\\
970612B & $ -0.89 $ & $ 37.71 $ & $ 14.53 \pm 2.49 $  &  $ 38.00 $\\
970625B & $ -1.35 $ & $ 48.98 $ & $ 178.88 \pm 9.79 $  &  $ 15.00 $\\
970627B & $ -0.73 $ & $ 15.86 $ & $ 100.03 \pm 7.58 $  &  $ 15.00 $\\
970706 & $ -9.01 $ & $ 72.25 $ & $ 15.69 \pm 0.61 $  &  $ 59.00 $\\
970816 & $ -0.06 $ & $ 6.61 $ & $ 51.43 \pm 2.74 $  &  $ 6.00 $\\
971027A & $ -1.66 $ & $ 12.20 $ & $ 26.89 \pm 1.46 $  &  $ 11.00 $\\
971223C & $ -6.22 $ & $ 50.18 $ & $ 52.58 \pm 4.13 $  &  $ 47.00 $\\
980203B & $ 0.38 $ & $ 48.75 $ & $ 217.07 \pm 8.81 $  &  $ 23.00 $\\
980306C & $ 0.62 $ & $ 28.25 $ & $ 79.07 \pm 1.95 $  &  $ 21.00 $\\
980329A & $ -1.06 $ & $ 36.93 $ & $ 73.26 \pm 4.15 $  &  $ 19.00 $\\
980428 & $ -5.05 $ & $ 88.46 $ & $ 21.72 \pm 1.34 $  &  $ 100.00 $\\
980615B & $ 0.94 $ & $ 97.48 $ & $ 85.10 \pm 5.01 $  &  $ 64.00 $\\
980827C & $ 0.33 $ & $ 87.24 $ & $ 158.30 \pm 5.36 $  &  $ 51.00 $\\
981111 & $ -6.39 $ & $ 48.81 $ & $ 35.91 \pm 2.67 $  &  $ 34.00 $\\
990128 & $ 0.67 $ & $ 11.30 $ & $ 121.88 \pm 3.11 $  &  $ 8.00 $\\
990620 & $ 0.42 $ & $ 13.97 $ & $ 38.68 \pm 1.72 $  &  $ 16.00 $\\
990705 & $ -0.23 $ & $ 41.19 $ & $ 63.92 \pm 3.93 $  &  $ 32.00 $\\
990913A & $ 0.03 $ & $ 44.54 $ & $ 183.03 \pm 8.08 $  &  $ 40.00 $\\
991124B & $ -1.65 $ & $ 25.31 $ & $ 8.01 \pm 0.62 $  &  $ 28.00 $\\
991216B & $ 0.46 $ & $ 25.42 $ & $ 416.88 \pm 11.96 $  &  $ 15.00 $\\
000115 & $ 0.04 $ & $ 25.71 $ & $ 200.84 \pm 8.42 $  &  $ 15.00 $\\
000214A & $ 0.37 $ & $ 8.75 $ & $ 58.66 \pm 3.60 $  &  $ 8.00 $\\
000218B & $ 0.26 $ & $ 23.70 $ & $ 258.43 \pm 11.67 $  &  $ 20.00 $\\
000419 & $ 0.72 $ & $ 20.70 $ & $ 21.65 \pm 0.82 $  &  $ 20.00 $\\
000630 & $ 0.94 $ & $ 44.55 $ & $ 21.76 \pm 2.02 $  &  $ 26.00 $\\
000718B & $ -0.19 $ & $ 97.05 $ & $ 67.51 \pm 2.98 $  &  $ 34.00 $\\
001004 & $ 1.10 $ & $ 11.20 $ & $ 191.46 \pm 8.26 $  &  $ 9.00 $\\
001011C & $ 0.94 $ & $ 31.62 $ & $ 29.67 \pm 1.42 $  &  $ 24.00 $\\
001212B & $ 0.64 $ & $ 72.46 $ & $ 45.83 \pm 3.10 $  &  $ 67.00 $\\
010109 & $ 0.90 $ & $ 22.17 $ & $ 293.48 \pm 6.62 $  &  $ 7.00 $\\
010317 & $ 0.87 $ & $ 31.03 $ & $ 210.87 \pm 8.65 $  &  $ 30.00 $\\
010408B & $ 0.23 $ & $ 6.40 $ & $ 199.33 \pm 8.39 $  &  $ 3.81 $\\
010412 & $ -1.49 $ & $ 65.48 $ & $ 24.62 \pm 2.47 $  &  $ 60.00 $\\
010504 & $ -0.12 $ & $ 19.84 $ & $ 42.79 \pm 4.07 $  &  $ 15.00 $\\
010710B & $ 1.06 $ & $ 27.05 $ & $ 53.73 \pm 4.50 $  &  $ 20.00 $\\
010922 & $ 0.60 $ & $ 41.52 $ & $ 19.20 \pm 1.30 $  &  $ 40.00 $\\
011003 & $ -0.94 $ & $ 45.41 $ & $ 36.72 \pm 1.81 $  &  $ 34.00 $\\
\enddata
\tablecomments{The PDS is calculated in the time interval reported.}
\tablenotetext{a}{Referred to the {\em BeppoSAX}/GRBM trigger time.}
\end{deluxetable}
\twocolumn

\onecolumn
\begin{deluxetable}{lrr}
\tablecolumns{3}
\tabletypesize{\scriptsize}
\tablecaption{White noise level (Leahy normalisation). {\em BeppoSAX} HR sample.
\label{tbl-4}}
\tablewidth{0pt}
\tablehead{
\colhead{GRB} & \colhead{$\langle P\rangle$ ($f>30$~Hz)} & \colhead{$\langle P\rangle$ ($f>50$~Hz)} }
\startdata
970111  & $ 2.13 \pm  0.09 $ & $ 2.11 \pm  0.15 $\\
970117B  & $ 2.11 \pm  0.14 $ & $ 2.17 \pm  0.22 $\\
970315A  & $ 2.09 \pm  0.13 $ & $ 2.09 \pm  0.20 $\\
970517B  & $ 2.50 \pm  0.31 $ & $ 2.19 \pm  0.46 $\\
970601  & $ 2.18 \pm  0.10 $ & $ 2.24 \pm  0.16 $\\
970612B  & $ 2.15 \pm  0.10 $ & $ 2.14 \pm  0.15 $\\
970625B  & $ 2.10 \pm  0.08 $ & $ 2.04 \pm  0.13 $\\
970627B  & $ 2.02 \pm  0.15 $ & $ 1.94 \pm  0.26 $\\
970706  & $ 2.09 \pm  0.07 $ & $ 2.13 \pm  0.10 $\\
970816  & $ 1.75 \pm  0.22 $ & $ 1.88 \pm  0.35 $\\
971027A  & $ 1.98 \pm  0.16 $ & $ 1.95 \pm  0.25 $\\
971223C  & $ 2.13 \pm  0.08 $ & $ 2.22 \pm  0.13 $\\
980203B  & $ 2.07 \pm  0.09 $ & $ 2.04 \pm  0.13 $\\
980306C  & $ 1.96 \pm  0.11 $ & $ 1.97 \pm  0.17 $\\
980329A  & $ 2.06 \pm  0.10 $ & $ 1.98 \pm  0.15 $\\
980428  & $ 2.08 \pm  0.06 $ & $ 2.14 \pm  0.10 $\\
980615B  & $ 2.10 \pm  0.06 $ & $ 2.08 \pm  0.09 $\\
980827C  & $ 2.08 \pm  0.06 $ & $ 2.18 \pm  0.10 $\\
981111  & $ 2.16 \pm  0.08 $ & $ 2.19 \pm  0.13 $\\
990128  & $ 2.11 \pm  0.19 $ & $ 2.08 \pm  0.29 $\\
990620  & $ 2.13 \pm  0.16 $ & $ 2.12 \pm  0.26 $\\
990705  & $ 2.07 \pm  0.09 $ & $ 1.97 \pm  0.14 $\\
990913A  & $ 1.94 \pm  0.09 $ & $ 1.95 \pm  0.14 $\\
991124B  & $ 2.11 \pm  0.12 $ & $ 2.14 \pm  0.18 $\\
991216B  & $ 2.01 \pm  0.12 $ & $ 2.01 \pm  0.19 $\\
000115  & $ 2.06 \pm  0.12 $ & $ 2.03 \pm  0.18 $\\
000214A  & $ 2.12 \pm  0.21 $ & $ 1.91 \pm  0.32 $\\
000218B  & $ 2.44 \pm  0.13 $ & $ 1.81 \pm  0.18 $\\
000419  & $ 2.12 \pm  0.13 $ & $ 2.05 \pm  0.21 $\\
000630  & $ 2.03 \pm  0.09 $ & $ 2.03 \pm  0.14 $\\
000718B  & $ 2.10 \pm  0.06 $ & $ 2.05 \pm  0.09 $\\
001004  & $ 2.05 \pm  0.19 $ & $ 1.99 \pm  0.29 $\\
001011C  & $ 2.07 \pm  0.11 $ & $ 2.00 \pm  0.18 $\\
001212B  & $ 2.04 \pm  0.07 $ & $ 2.12 \pm  0.11 $\\
010109  & $ 1.93 \pm  0.13 $ & $ 1.87 \pm  0.20 $\\
010317  & $ 2.29 \pm  0.11 $ & $ 2.05 \pm  0.17 $\\
010408B  & $ 1.76 \pm  0.23 $ & $ 1.4 \pm  0.34 $\\
010412  & $ 2.01 \pm  0.07 $ & $ 2.07 \pm  0.11 $\\
010504  & $ 2.03 \pm  0.14 $ & $ 1.98 \pm  0.25 $\\
010710B  & $ 2.07 \pm  0.12 $ & $ 2.00 \pm  0.18 $\\
010922  & $ 2.06 \pm  0.09 $ & $ 2.09 \pm  0.15 $\\
011003  & $ 2.10 \pm  0.09 $ & $ 2.17 \pm  0.14 $\\
\tablecomments{Table of white noise level at $f>30$~Hz and at $f>50$~Hz related to the sub-sample of 42 GRBs detected by {\em BeppoSAX}/GRBM with 7.8 ms time resolution. Uncertainties at 1$\sigma$}
\enddata
\end{deluxetable}
\twocolumn

\onecolumn
\begin{deluxetable}{lrr}
\tablecolumns{3}
\tabletypesize{\scriptsize}
\tablecaption{White noise level (Leahy normalisation). {\em BeppoSAX} VHR sample.
\label{tbl-5}}
\tablewidth{0pt}
\tablehead{
\colhead{GRB} & \colhead{$\langle P\rangle$ ($f>30$~Hz)} & \colhead{$\langle P\rangle$ ($f>50$~Hz)} }
\startdata
970315A  & $ 2.00 \pm  0.03 $ & $ 2.00 \pm  0.03 $\\
970517B  & $ 2.03 \pm  0.03 $ & $ 2.03 \pm  0.03 $\\
970601  & $ 2.12 \pm  0.03 $ & $ 2.10 \pm  0.04 $\\
970625B  & $ 2.02 \pm  0.03 $ & $ 2.01 \pm  0.03 $\\
970627B  & $ 2.02 \pm  0.03 $ & $ 2.02 \pm  0.03 $\\
970816  & $ 1.97 \pm  0.04 $ & $ 1.98 \pm  0.04 $\\
980203B  & $ 2.01 \pm  0.03 $ & $ 2.01 \pm  0.03 $\\
990128  & $ 2.04 \pm  0.03 $ & $ 2.04 \pm  0.03 $\\
990620  & $ 2.06 \pm  0.03 $ & $ 2.06 \pm  0.03 $\\
990705  & $ 2.04 \pm  0.03 $ & $ 2.04 \pm  0.03 $\\
990913A  & $ 1.97 \pm  0.03 $ & $ 1.97 \pm  0.03 $\\
991216B  & $ 1.91 \pm  0.03 $ & $ 1.91 \pm  0.03 $\\
000115  & $ 2.01 \pm  0.03 $ & $ 2.01 \pm  0.03 $\\
000214A  & $ 2.03 \pm  0.03 $ & $ 2.03 \pm  0.03 $\\
000630  & $ 2.05 \pm  0.03 $ & $ 2.05 \pm  0.03 $\\
001004  & $ 1.99 \pm  0.03 $ & $ 1.98 \pm  0.03 $\\
001212B  & $ 2.06 \pm  0.03 $ & $ 2.06 \pm  0.03 $\\
010109  & $ 1.94 \pm  0.03 $ & $ 1.94 \pm  0.03 $\\
010317  & $ 2.04 \pm  0.03 $ & $ 2.03 \pm  0.03 $\\
010408B  & $ 1.99 \pm  0.03 $ & $ 1.99 \pm  0.03 $\\
010504  & $ 2.01 \pm  0.04 $ & $ 2.01 \pm  0.04 $\\
\tablecomments{Table of white noise level at $f>30$~Hz and at $f>50$~Hz related to the sub sample of 25 GRBs detected by {\em BeppoSAX}/GRBM with 0.5 ms time resolution. Uncertainties at 1$\sigma$}
\enddata
\end{deluxetable}
\twocolumn

\onecolumn
\begin{deluxetable}{lrcccccccc}
\tablecolumns{10}
\tabletypesize{\scriptsize}
\rotate
\tablecaption{Best fit parameters of the average PDS for different samples of GRBs
\label{tbl-6}}
\tablewidth{0pt}
\tablehead{\colhead{Sample} & \colhead{Size} & \colhead{Norm}  &
\colhead{$\alpha_1$} & \colhead{$f_{\rm b}$} & \colhead{$\alpha_2$} & \colhead{} &
\colhead{$f_{\rm b2}$} & \colhead{$\alpha_3$} & \colhead{$\chi^2$/dof}\\
 & & & & \colhead{($10^{-2}$~Hz)} & & & \colhead{(Hz)} & &}
\startdata
{\em Fermi}/GBM (8--1000 keV)\tablenotemark{a} & 205 & $5.0_{-0.9}^{+1.2}$ & $1.06_{-0.07}^{+0.05}$ & $ 5.5_{-0.7}^{+0.8}$ & $1.75_{-0.03}^{+0.03}$ & 
                & --                  & --                  & $110/100$\\
{\em Fermi}/GBM (8--40 keV)\tablenotemark{a} & 155 & $3.9_{-1.1}^{+1.5}$ & $1.20_{-0.08}^{+0.07}$ & $ 6.4_{-1.2}^{+1.4}$ & $1.95_{-0.06}^{+0.07}$ & 
                & --                  & --                  & $78/54$\\
{\em Fermi}/GBM (40--200 keV)\tablenotemark{a} & 201 & $5.1_{-1.1}^{+0.7}$ & $1.03_{-0.04}^{+0.06}$ & $ 5.5_{-0.5}^{+1.0}$ & $1.67_{-0.03}^{+0.02}$ & 
                & --                  & --                  & $130/115$\\
{\em Fermi}/GBM (200--1000 keV)\tablenotemark{a} & 74 & $7.3_{-4.2}^{+5.8}$ & $1.05_{-0.09}^{+0.08}$ & $ 3.8_{-1.5}^{+3.4}$ & $1.47_{-0.04}^{+0.06}$ & 
                & --                  & --                  & $79/72$\\
{\em Fermi}/GBM FRED (8--1000 keV)\tablenotemark{a} & 10 & $3.8_{-1.9}^{+3.0}$ & $1.32_{-0.10}^{+0.10}$ & $ 6.3_{-1.9}^{+3.1}$ & $2.53_{-0.24}^{+0.39}$ & 
                & --                  & --                  & $16/14$\\
{\em BeppoSAX}/GRBM HR (40--700 keV)\tablenotemark{b} & 42 & $0.021_{-0.006}^{+0.011}$ & --                  & --                   & $1.49_{-0.04}^{+0.04}$ & 
                & $1.9_{-0.4}^{+0.4}$ & $2.46_{-0.31}^{+0.44}$ & $145/143$\\
{\em BeppoSAX}/GRBM VHR (40--700 keV)\tablenotemark{b} & 25 & $0.040_{-0.022}^{+0.048}$ & --                  & --                   & $1.52_{-0.17}^{+0.17}$ & 
                & $2.6_{-0.9}^{+1.0}$ & $2.91_{-0.41}^{+0.51}$ & $4/7$\\
{\em BeppoSAX}/GRBM HR+VHR (40--700 keV)\tablenotemark{b,c} & 42+25 & $0.016_{-0.005}^{+0.006}$ ; $0.053_{-0.014}^{+0.017}$ & --                  & --                   & $1.50_{-0.04}^{+0.03}$ & 
                & $2.1_{-0.3}^{+0.4}$ & $2.69_{-0.20}^{+0.27}$ & $165/161$\\
{\em Fermi}/GBM VHR (40--700 keV)\tablenotemark{b,d} & 96 & $0.029_{-0.011}^{+0.015}$ & --                  & --                   & $1.65_{-0.03}^{+0.03}$ & 
                & $1.1_{-0.2}^{+0.3}$ & $2.41_{-0.19}^{+0.34}$ & $213/200$\\
{\em BeppoSAX}/GRBM HR+VHR + {\em Fermi}/GBM VHR (40--700 keV))\tablenotemark{b,e} & 42+25+96 & $0.027_{-0.008}^{+0.014}$ ; $0.088_{-0.025}^{+0.042}$ ; $0.019_{-0.006}^{+0.010}$ & --                  & --                   & $1.60_{-0.03}^{+0.02}$ & 
                & $1.4_{-0.3}^{+0.3}$ & $2.33_{-0.13}^{+0.15}$ & $502/365$\\

{\em Fermi}/GBM (15--150 keV)\tablenotemark{a} & 200 & $5.1_{-1.0}^{+1.2}$ & $1.06_{-0.07}^{+0.06}$ & $ 5.5_{-0.7}^{+0.9}$ & $1.78_{-0.03}^{+0.04}$ & 
                & --                  & --                  & $95/91$\\
\enddata
\tablecomments{Best--fitting parameters of the average PDS of each sample within different energy bands ({\em Fermi}) and time resolution ({\em BeppoSAX}).}
\tablenotetext{a}{Low frequency break}
\tablenotetext{b}{High frequency break}
\tablenotetext{c}{Joint fitting of two samples with different time resolutions obtained through the minimization of the joint likelihood.
The normalisation parameters refer to $7.8$ and $0.5$--ms time resolution, respectively.}
\tablenotetext{d}{In this case, the best--fitting parameters were found by fitting the average spectra in the same frequency range considered for
{\em BeppoSAX} from $0.02$ to 1000~Hz.}
\tablenotetext{e}{Joint fitting of three samples with different time resolutions obtained through the minimization of the joint likelihood.
The normalisation parameters refer to $7.8$ and $0.5$ ms time resolution for {\em BeppoSAX} and $0.5$ ms for the {\em Fermi}, respectively.}
\end{deluxetable}
\twocolumn

\end{document}